\input harvmac
\input mssymb
\input labeldefs.tmp
\writedefs
\overfullrule=0pt
%
\input epsf
\newdimen\xfigunit
\global\xfigunit=4144sp
\def\figmag{1}
\newdimen\x \newdimen\y
\def\pic(#1,#2)(#3,#4)#5{%
\global\x=-#3\xfigunit%
\global\advance\x by-#1\xfigunit%
\global\y=-#4\xfigunit%
\def\epsfsize##1##2{\figmag##1}%
\epsfbox{#5}%
}
\def\put(#1,#2)#3{{\advance\x by#1\xfigunit\advance\y by#2\xfigunit%
\x=\figmag\x%
\y=\figmag\y%
\rlap{\kern\x%
\raise\y\hbox{#3}}}}
\def\fig#1#2#3{
\xdef#1{\the\figno}
\writedef{#1\leftbracket \the\figno}
\nobreak
\par\begingroup\parindent=0pt\leftskip=1cm\rightskip=1cm\parindent=0pt
\baselineskip=11pt
\midinsert
\centerline{#3}
\vskip 12pt
{\bf Fig.\ \the\figno:} #2\par
\endinsert\endgroup\par
\goodbreak
\global\advance\figno by1
}
\global\newcount\tabno \global\tabno=1
\def\tab#1#2#3{
\xdef#1{\the\tabno}
\writedef{#1\leftbracket \the\tabno}
\nobreak
\par\begingroup\parindent=0pt\leftskip=1cm\rightskip=1cm\parindent=0pt
\baselineskip=11pt
\midinsert
\centerline{#3}
\vskip 12pt
{\bf Tab.\ \the\tabno:} #2\par
\endinsert\endgroup\par
\goodbreak
\global\advance\tabno by1
}
\def\der{\partial}
\def\d{{\rm d}}
\def\e#1{{\rm e}^{#1}}
\def\E#1{{\rm e}^{\textstyle #1}}%

\def\braket#1#2{\big< #1 \big| #2 \big>}

\def\A{{\rm A}}
\def\B{{\rm B}}
\def\M{{\rm M}}
\def\U{{\rm U}}
\def\V{{\rm V}}
\def\frac#1#2{{\textstyle{#1\over#2}}}
%
\def\pre#1{ (preprint {\tt #1})}
%
%
%
%
\lref\DKK{J.M. Daul, V. Kazakov, I. Kostov, {\it Nucl. Phys.} B 409 (1993), 
311\pre{hep-th/9303093}.}
%
\lref\UT{K.~Ueno and K.~Takasaki, {\it Adv. Stud. Pure Math.} 4,
editors H.~Morikawa and K.~Okamoto (1984), 1.}
%
\lref\MWZ{A.~Marshakov, P.~Wiegmann, A.~Zabrodin,
{\sl Integrable Structure of the Dirichlet Boundary Problem in Two 
Dimensions}\pre{hep-th/0109048}.}
%
\lref\WZ{P.~Wiegmann, A.~Zabrodin, {\sl Conformal Maps and Integrable
Hierarchies}\pre{hep-th/9909147}.}
\lref\TT{K.~Takasaki, T.~Takebe, {\it Lett. Math. Phys.} 28 (1993), 
165--176\pre{hep-th/9301070}.}
%
\lref\TTb{K.~Takasaki, T.~Takebe, {\it Rev. Math. Phys.} 7 (1995), 
743--808\pre{hep-th/9405096}.}
%
\lref\Kri{I.M.~Krichever, {\it Funct. Anal. Appl.} 22 (1989), 200--213\semi
{\it Commun. Pure Appl. Math.} 47 (1992), 437\pre{hep-th/9205110}.}
%
\lref\HC{
Harish~Chandra, {\it Amer. J. Math.} 79 (1957) 87--120.}
\lref\IZ{
C.~Itzykson and J.-B.~Zuber, {\it J. Math. Phys.} 21 (1980) 411.}
\lref\Meh{M.L.~Mehta, {\it Commun. Math. Phys.} 79 (1981), 327.}
%
\lref\PZJ{P.~Zinn-Justin, {\it Commun. Math. Phys.} 194
(1998), 631\pre{cond-mat/9705044}.}
%
\lref\MAT{A.~Matytsin, {\it Nucl. Phys.} B 411 (1994) 805.}
%
\lref\PZJb{P.~Zinn-Justin, {\it Phys. Rev.} E59 (1999), 4884\pre{math-ph/9810010}.}
%
\lref\BIPZ{E.~Br\'ezin, C.~Itzykson, G.~Parisi and J.-B.~Zuber,
{\it Commun. Math. Phys.} 59 (1978), 35.}
\lref\Sha{S.~L.~Shatashvili, {\it Commun. Math. Phys.} 154 (1993), 
421\pre{hep-th/9209083}.}
%
\lref\PZJc{P.~Zinn-Justin,
{\it Phys. Rev.} E 62 (2000), 3411\pre{math-ph/0005008}.}
%
\lref\AvM{M.~Adler, P.~van Moerbeke, {\it Commun. Pure Appl. Math.}
50 (1997), 241\pre{hep-th/9706182}.}
%
%
\lref\KAZBOUL{V.~Kazakov, {\it Phys. Lett.} A119 (1986), 140\semi
D.V.~Boulatov and V.~Kazakov, {\it Phys. Lett.} B186 (1987), 379.}
\lref\DFGZJ{P. Di Francesco, P. Ginsparg and J. Zinn-Justin, 
{\it Phys. Rep.} 254 (1995), 1--133.}
\lref\GZ{A.~Guionnet, O.~Zeitouni, {\sl Large deviations asymptotics
for spherical integrals}, preprint.}
%
\lref\Co{B.~Collins, {\sl Moments and Cumulants of Polynomial random
variables on unitary groups, the Itzykson--Zuber integral and free
probability}\pre{math-ph/0205010}.}
\lref\AI{D.~Altschuler, C.~Itzykson, {\it Ann. Inst. Henri Poincar\'e
(phys. th\'eor.)} 54 (1991), 1.}
%
\lref\Sto{M.~Stone, {\it Nucl. Phys.} B 314 (1989), 557.}
%
\lref\Tut{W.T.~Tutte, {\it Canad. J. Math.} 14 (1962), 402--417.}
%
\lref\Spe{R.~Speicher, {\it Math. Ann.} 298 (1994), 611.}
%
\lref\VDN{D.V.~Voiculescu, K.J.~Dykema, A.~Nica, {\it Free Random Variables}
(1992), American Mathematical Society.}
%
\lref\Kos{I.~Kostov, {\sl String Equation for String Theory on a 
Circle}\pre{hep-th/0107247}.}
%
\Title{
\vbox{\baselineskip12pt\hbox{{\tt math-ph/0202045}}}}
{{\vbox {
\vskip-10mm
\centerline{HCIZ integral}
\vskip2pt
\centerline{and 2D Toda lattice hierarchy}
}}}
\medskip
\centerline{P.~Zinn-Justin}\medskip
\centerline{\sl Laboratoire de Physique Th\'eorique et Mod\`eles Statistiques}
\centerline{\sl Universit\'e Paris-Sud, B\^atiment 100}
\centerline{\sl 91405 Orsay Cedex, France}
\vskip .2in
\noindent
The expression of the large $N$
Harish Chandra--Itzykson--Zuber (HCIZ) integral in terms of the moments
of the two matrices is investigated using an auxiliary unitary 
two-matrix model,
the associated biorthogonal polynomials and integrable hierarchy. We find
that the large $N$ HCIZ integral is governed by the dispersionless
Toda lattice hierarchy and derive its string equation.
We use this to obtain various exact results on its expansion in powers
of the moments.
\Date{02/2002}

\newsec{Introduction and definitions}
The Harish Chandra--Itzykson--Zuber integral \refs{\HC,\IZ} 
plays a key role in matrix models \refs{\IZ,\DKK,\Meh,\KAZBOUL,\DFGZJ},
in particular in relation to (discretized)
two-dimensional quantum gravity. It also
displays interesting connections \refs{\PZJb,\Co}
with free probability theory \refs{\VDN,\Spe}.
It is defined by
\eqn\HCIZ{
I_N(\A,\B)=\int_{\Omega\in U(N)} \d\Omega\, \E{N \tr \A\Omega \B\Omega^\dagger}
}
where $\A$ and $\B$ are $N\times N$ matrices and $\d\Omega$
is the normalized Haar measure on $U(N)$. There are many ways
to evaluate this integral \refs{\HC,\IZ,\Meh,\Sto,\AI,\Sha}; the result
depends only on the eigenvalues $a_i$ of $\A$ and $b_i$ of $\B$,
and is given by
\eqn\HCIZb{
I_N(\A,\B)={\det(\e{N a_i b_j})\over\Delta(a_i)\Delta(b_j)}
}
Here $\Delta(\cdot)$ denotes
the Vandermonde determinant: $\Delta(a_i)=\prod_{i<j} (a_i-a_j)$.
In all equations, constants that only depend on $N$ will be ignored; they
can be restored when required.

For practical applications it is important to look at the large $N$ limit
of the HCIZ integral: it corresponds
to the ``planar'' limit
of matrix models. One considers a sequence of $N\times N$ matrices
$\A$ and $\B$, whose spectral distributions converge to some fixed
distributions as $N$ goes to infinity. It is known 
(and in fact, proved rigorously \GZ) that the following quantity,
the ``large $N$ free energy'', is well-defined:
\eqn\defF{
F[\theta_q,\tilde{\theta}_q]=\lim_{N\to\infty} {\log I_N(\A,\B)\over N^2}
}
and depends on the spectral distributions of $\A$ and $\B$ 
only through their moments:
\eqn\deftq{
\theta_q=\lim_{N\to\infty}{\tr\over N} \A^q\qquad
\tilde{\theta}_q=\lim_{N\to\infty}{\tr\over N} \B^q\qquad q\ge 1
}
We also define for future use the generating series of these moments,
that is the resolvents of $\A$ and $\B$:
\eqna\defres
$$\eqalignno{
G(a)&\equiv\lim_{N\to\infty}{\tr\over N} {1\over a-\A}
= 1/a + \sum_{q\ge 1} \theta_q / a^{q+1}&\defres{a}\cr
\tilde{G}(b)&\equiv \lim_{N\to\infty}{\tr\over N} {1\over b-\B}
= 1/b+\sum_{q\ge 1} \tilde{\theta}_q / b^{q+1}&\defres{b}\cr
}$$

As a function of the $\theta_q$ and the $\tilde{\theta}_q$, 
$F$ has a power series expansion
around $0$. We shall investigate here the determination of the coefficients
of this expansion, as well as the more general question of the evolution
of $F$ as one varies the $\theta_q$ and $\tilde{\theta}_q$. 
We shall see that this
evolution is governed by an infinite set of partial differential equations,
a particular scaling of the Toda lattice hierarchy.

More precisely, we shall rewrite in section 2 the HCIZ integral as a unitary
two-matrix model, then show in section 3 that the latter satisfies the
Toda lattice hierarchy; in section 4, we shall take the large $N$ limit
which relates $F$ to the dispersionless Toda lattice hierarchy; finally,
we shall apply this technology to the expansion of $F$ in section 5.
Section 6 is devoted to a summary of results.

\newsec{HCIZ integral as a unitary two-matrix model}
We shall first use a trick to rewrite the HCIZ integral as an integral
over two matrices. The derivation can be performed either by computing residues
in the two-matrix integral, or by character expansion; we choose the latter.
We thus start with the following identity, valid for any $N\times N$ matrix $\M$:
\eqn\id{
\E{N\tr \M}=\sum_{R\ge 0} c_R d_R \chi_R(\M)
}
Let us explain the symbols used. $R$ stands for an analytic irrep of 
$GL(N)$ with positive
highest weights, or the associated Young diagram;
more explicitly, it can be described by its ``shifted''
highest weights $h_i$, $1\le i\le N$ which form a strictly decreasing sequence
(they are related to the usual highest weights $m_i$ by $h_i=m_i+N-i$).
$\chi_R(\M)$ is the character associated to the irrep $R$ and the matrix $\M$.
$c_R$ and $d_R$ are coefficient of the expansion:
\eqn\defcd{
c_R={N^{|R|}\over\prod_{i=1}^N h_i!}\qquad d_R=\Delta(h_i)
}
where $|R|=\sum_i h_i - N(N-1)/2$ is the number of boxes of $R$.
These explicit expressions of the coefficients
play no role in what follows since the trick we use
is quite general.\foot{One can treat similarly
any expression of the form ${\det f(a_i,b_j)\over \Delta(a_i)\Delta(b_j)}$,
cf Eq.~\HCIZb. However here we do not use Eq~\HCIZb\ -- in fact,
we prove it along the way.}

Applying the identity \id\ to the definition of the HCIZ integral \HCIZ\ 
and using orthogonality relations for matrix elements of irreps yields 
\eqn\HCIZc{
I_N(\A,\B)=\sum_{R\ge 0} c_R \chi_R(\A) \chi_R(\B)
}
As a side remark, we note that using Weyl's formula 
for characters ($\chi_R(\A)=\det(a_i^{h_j})/\Delta(a_i)$),
recombining the
two resulting determinants into one single determinant and performing
the summation, one recovers Eq.~\HCIZb.

We now want to compare expression \HCIZc\ with the partition function of
a matrix model of two $N\times N$ matrices $\U$ and $\V$:
\eqn\twoMM{
\tau_N[t_q,\tilde{t}_q]=\oint\!\!\oint \d \U \d \V
\E{\sum_{q\ge 1} t_q \tr \U^q+\sum_{q\ge 1} \tilde{t}_q \tr \V^q + \tr \U^{-1} \V^{-1}}
}
Here the $t_q$ are as yet unknown coefficients; they must however satisfy
$\sup_q |t_q|^{1/q}<m<\infty$ and 
$\sup_q |\tilde{t}_q|^{1/q}<\tilde{m}<\infty$;
this renders the summation convergent provided that the spectral radii
of $\U$ and $\V$ are less or equal to $1/m$, $1/\tilde{m}$. 
Since the integrand is analytic in $\U$
and $\V$ the exact contours of integration are irrelevant; 
for the sake of definiteness, we choose
$\U$ and $\V$ to be unitary up to a multiplicative constant:
$\U\U^\dagger=m^{-2}$, $\V\V^\dagger=\tilde{m}^{-2}$.
The measure of integration is simply the normalized
Haar measures for $m\, \U$ and $\tilde{m}\, \V$. All these details can
be safely ignored if one remembers that in the integral \twoMM\ only
the contribution coming from poles at the origin should be picked up.

Note that we have defined \twoMM\ so that $N$ does not appear anywhere
explicitly in the action. It is however convenient to rescale the matrices to
restore the usual $N$ dependence of matrix models:
\eqn\twoMMb{
\tau_N[t_q,\tilde{t}_q]=
\oint\!\!\oint \d \U \d \V
\E{\sum_{q\ge 1} t_q N^{-q/2} \tr \U^q+\sum_{q\ge 1} \tilde{t}_q N^{-q/2}
\tr \V^q + N\tr \U^{-1} \V^{-1}}
}
Next we apply one of the forms of the Cauchy identity:
\eqn\cau{
\E{\sum_{q\ge 1} t_q N^{-q/2} \tr \U^q}=\sum_{R\ge 0}
s_R(t_q N^{-q/2}) \chi_R(\U)
}
where $s_R$ is the Schur function associated to the Young diagram $R$.

Plugging twice \cau\ as well as the identity \id\ into \twoMMb\ results in:
\eqn\twoMMc{
\tau_N[t_q,\tilde{t}_q]=\sum_{R,R_1,R_2\ge 0} s_{R_1}(t_q N^{-q/2})
s_{R_2}(\tilde{t}_q N^{-q/2}) c_R d_R \oint\!\!\oint \d \U \d \V
\chi_{R_1}(\U) \chi_{R_2}(\V) \chi_R(\U^{-1}\V^{-1})
}
Again, orthogonality relations imply that
\eqn\twoMMd{
\tau_N[t_q,\tilde{t}_q]=\sum_{R\ge 0\atop \# {\rm rows} \le N} c_R 
s_R(t_q N^{-q/2}) s_R(\tilde{t}_q N^{-q/2})
}
Note that we have written explicitly the constraint that
the Young diagram $R$ should have no more than $N$ rows, since
contrary to prior expressions the summand in Eq.~\twoMMd\ is
not necessarily zero if the number of rows exceeds $N$.

Comparing Eqs.~\twoMMd\ and \HCIZc, we see that the two expressions are
equal (up to a numerical factor) iff
\eqn\iff{
t_q N^{-q/2} = {1\over q} \tr \A^q\qquad
\tilde{t}_q N^{-q/2} = {1\over q} \tr \B^q
}
or in terms of the normalized moments introduced before
\eqn\iffb{
t_q=N^{q/2+1} {\theta_q\over q}\qquad \tilde{t}_q=N^{q/2+1}
{\tilde{\theta}_q\over q}
}
The constraint on the number of rows of $R$ is automatically satisfied
if the $t_q$ are of the form \iff. However, the expression \twoMMd\ has the
advantage that it is defined for arbitrary (independent) $t_q$ at finite $N$,
which is not the case of the original HCIZ integral (where there are only
$N$ independent $t_q$). Also note that the condition of finiteness of
$\sup_{q\ge 1} 
|t_q|^{1/q}$ and $\sup_{q\ge 1} |\tilde{t}_q|^{1/q}$ is automatically
satisfied at finite $N$ if $t_q$ is of the form \iff, and remains
true as $N\to\infty$ provided the spectra of $\A$ and $\B$ remain bounded.

\newsec{Biorthogonal polynomials and Toda lattice hierarchy}
We have found in the previous section that calculating the HCIZ integral is
equivalent to solving the matrix model given by Eq.~\twoMM. A convenient way to
do so, at least formally, is via biorthogonal polynomials. This is standard
material, and is very similar to derivations found in the literature
\AvM. $\tau_N$ will then be identified with a tau function of the 2D Toda
lattice hierarchy.

\subsec{Setup}
We start with the following determinant form of the partition function \twoMM,
obtained by diagonalizing $\U$ and $\V$:
\eqn\detform{
\tau_N=\det\left(\oint\!\!\oint {\d u\over 2\pi i u}{\d v\over 2\pi i v}
u^j v^i \e{\sum_{q\ge 1} t_q u^q+\sum_{q\ge 1} \tilde{t}_q v^q+u^{-1}v^{-1}}
\right)_{0\le i,j\le N-1}
}
where once again the contours of integrations are sufficiently small
circles around the origin.

This suggests to introduce a
non-degenerate bilinear form on the space of polynomials by
\eqn\bil{
\braket{q}{p}=
\oint\!\!\oint {\d u\over 2 \pi i u}{\d v\over 2 \pi i v}
p(u) q(v)
\e{\sum_{q\ge 1} t_q u^q+\sum_{q\ge 1} \tilde{t}_q v^q+u^{-1}v^{-1}}
}
so that $\tau_N=\det M_N$, with $M_N=(m_{ij})$, $0\le i,j\le N-1$ and
$m_{ij}=\braket{v^i}{u^j}$.

Next we define normalized biorthogonal polynomials $q_n(v)$ and $p_n(u)$ 
with respect to the bilinear form above on ${\Bbb C}[v]\times{\Bbb C}[u]$,
that is $\braket{q_n}{p_m}=\delta_{nm}$.
We write:
\eqn\biorth{p_n(u)=\sum_{k=0}^n p_{kn} u^k\qquad
q_n(v)=\sum_{k=0}^n q_{kn} v^k}
There is an arbitrariness in the normalization which is fixed
by assuming $p_{nn}=q_{nn}\equiv h_n^{-1}$.

Define upper triangular 
matrices $P_N=(p_{kn})$ and $Q_N=(q_{kn})$, $0\le n,k\le N-1$;
and $\tilde{P}_N={Q_N^T}^{-1}$. Orthonormality of these polynomials is
equivalent to writing
\eqn\orth{
Q_N^T M_N P_N=1
}
or $M_N=\tilde{P}_N P_N^{-1}$, with $P_N^{-1}$
upper triangular and $\tilde{P}_N$ lower 
triangular. In particular $\tau_N=\det M_N=\prod_{i=0}^{N-1} h_i^2$.

Next, introduce the semi-infinite matrices
$M=M_\infty$, $P=P_\infty$, $Q=Q_\infty$, $\tilde{P}=\tilde{P}_\infty$
and the shift matrix $Z$: $Z_{ij}=\delta_{i\,j+1}$. Set
\eqn\defuv{
U=P^{-1}Z P\qquad V=Q^{-1}Z Q
}
$U$ (resp.\ $V$) is the matrix of multiplication by $u$ (resp.\ $v$) in
the basis $(p_n(u))$ of ${\Bbb C}[u]$ (resp.\ $(q_n(v))$ of ${\Bbb C}[v]$).
Also set $\tilde{U}\equiv V^T=\tilde{P}^{-1} Z^T \tilde{P}$.
By definition of the $m_{ij}$, we have
\eqn\derM{
{\der M\over\der t_q} = M Z^q\qquad
\qquad
{\der M\over\der \tilde{t}_q} = {Z^T}^q M
}
We then easily find, using $M=\tilde{P} P^{-1}$,
\eqn\derP{
-P^{-1}{\der P\over\der t_q} + \tilde{P}^{-1} {\der \tilde{P}\over\der t_q}
=U^q
\qquad
-P^{-1}{\der P\over\der \tilde{t_q}}
+\tilde{P}^{-1}{\der\tilde{P}\over\der \tilde{t}_q} 
=\tilde{U}^q
}
$P^{-1}{\der P\over\der t_q}$ is upper triangular, whereas
$\tilde{P}^{-1}{\der\tilde{P}\over\der t_q}$ is lower triangular, and
they have opposite diagonal elements (and similarly
for derivatives with respect to $\tilde{t}_q$); so that if one defines
symbols
$(\cdots)_+=(\cdots)_{>0}+{1\over2}(\cdots)_0$ and
$(\cdots)_-=(\cdots)_{<0}+{1\over2}(\cdots)_0$ for lower and upper diagonal
parts respectively
(it is convenient to include one half of the diagonal part, though this does
not make them projections), we have
\eqna\linsys
$$\eqalignno{
{\der P\over\der t_q}=-P\,(U^q)_-\qquad&\qquad
{\der P\over\der \tilde{t}_q}=-P\,(\tilde{U}^q)_-
&\linsys{a}\cr
{\der \tilde{P}\over\der t_q}=\tilde{P}\, (U^q)_+
\qquad&\qquad
{\der \tilde{P}\over\der \tilde{t_q}}=\tilde{P}\,(\tilde{U}^q)_+
&\linsys{b}\cr
}$$
and finally
\eqna\lax
$$\eqalignno{
{\der U\over\der t_q}=-[(U^q)_+,U]\qquad&\qquad
{\der U\over\der \tilde{t}_q}=[(\tilde{U}^q)_-,U]
&\lax{a}\cr
{\der \tilde{U}\over\der t_q}=-[(U^q)_+,\tilde{U}]
\qquad&\qquad
{\der \tilde{U}\over\der \tilde{t_q}}=[(\tilde{U}^q)_-,\tilde{U}]&\lax{b}\cr
}$$
Equations \lax{} are equivalent to equations of Zakharov--Shabat type \UT
\eqna\ZS
$$\eqalignno{
{\der\over\der t_q}(U^r)_+ - {\der\over\der t_r} (U^q)_+
+[(U^q)_+,(U^r)_+]&=0&\ZS{a}\cr
{\der\over\der \tilde{t}_q}(U^r)_+ + {\der\over\der t_r} (\tilde{U}^q)_-
-[(\tilde{U}^q)_-,(U^r)_+]&=0&\ZS{b}\cr
{\der\over\der \tilde{t}_q}(\tilde{U}^r)_- 
- {\der\over\der \tilde{t}_r} (\tilde{U}^q)_-
-[(\tilde{U}^q)_-,(\tilde{U}^r)_-]&=0&\ZS{c}\cr
}
$$
which ensure compatibility of Eqs.~\linsys{}.
If we expand $U$ and $\tilde{U}$ in the following way:
\eqna\expu
$$\eqalignno{
U&=r \,Z+\sum_{k=0}^\infty u_k\, Z^T{}^k&\expu{a}\cr
\tilde{U}&=Z^T\, r+\sum_{k=0}^\infty Z^k\, v_k&\expu{b}\cr
}$$
where $r={\rm diag}(r_n)$, the $u_k={\rm diag}(u_{k;n})$ 
and the $v_k={\rm diag}(v_{k;n})$
are diagonal matrices, then Eqs.~\ZS{} are
the (semi-infinite) Toda lattice hierarchy of differential equations
for the coefficients of $U$ and $\tilde{U}$.

\subsec{Examples}
The simplest equation, the usual Toda lattice equation, which
is the case $q=r=1$ in Eq.~\ZS{b}, can be obtained directly
from the determinant form \detform\ by applying Jacobi's determinant identity,
which yields $\tau_{n+1}\tau_{n-1}=\tau_n{\der^2\over\der t_1\der \tilde{t}_1} 
\tau_n- {\der\over\der t_1}\tau_n{\der\over\der \tilde{t}_1} \tau_n$, or
\eqn\toda{
{\tau_{n+1}\tau_{n-1}\over \tau_n^2}={\der^2\over\der t_1\der \tilde{t}_1} 
\log \tau_n
}
The left hand side is nothing but $(h_n/h_{n-1})^2=r_n^2$.
This implies
\eqn\todab{
r_{n+1}^2+r_{n-1}^2-2 r_n^2 = {\der^2\over\der t_1\der \tilde{t}_1}
\log r_n^2
}
which is the Toda lattice equation in terms of the $\log r_n^2$.
In the next section we shall see how to use
this equation in the large $N$ limit.

Another equation one can derive (case $q=2$, $r=3$ of Eq.~\ZS{a})
is the usual Kadomtsev--Petviashvili
(KP) equation, 
which is given in terms of $\chi=2 {\der^2\over\der t_1^2}\log \tau_N$:
\eqn\KP{
3 {\der^2 \chi\over\der t_2^2}+{\der\over\der t_1} \left(-4 {\der \chi\over\der t_3}
+6 \chi {\der \chi\over \der t_1}+{\der^3 \chi\over \der t_1^3}\right)=0
}

\subsec{String equation}
So far we have not used the specific form of the interaction between
our two matrices in \twoMM. This occurs when one tries to determine
the {\it string equation}, that is an additional relation which fixes
a particular solution of the Toda hierarchy. Here it is very easy to
derive the string equation 
by arguments similar to those used in the standard two-matrix model.
We write that
\eqn\prese{
0=\oint\!\!\oint {\d u\over 2 \pi i u}{\d v\over 2 \pi i v}
u{\d\over\d u}\left[
p_j(u) q_i(v)
\e{\sum_{q\ge 1} t_q u^q+\sum_{q\ge 1} \tilde{t}_q v^q+u^{-1}v^{-1}}\right]
}
expand and obtain
\eqn\preseb{
UD+
\sum_{q\ge 1} q t_q U^{q}- \tilde{U}^{-1} U^{-1} =0
}
where $D$ is the derivative with respect to $u$ in the basis $p_n(u)$,
and $U^{-1}$ is a particular left inverse of $U$ whose matrix elements
are defined by Eq.~\prese, and similarly $\tilde{U}^{-1}$ is a particular
right inverse of $\tilde{U}$.
Using $[D,U]=1$ and $U^{-1}U=1$, 
we derive from Eq.~\preseb\ the string equation:
\eqn\se{
[U^{-1},\tilde{U}^{-1}]=1
}
Let us note that Eq.~\se\ is 
{\it not}\/ the usual string equation of the hermitean two-matrix model.

\newsec{Large $N$ limit}
\subsec{Classical/dispersionless limit}
Next we want to consider the large $N$ limit. As always in matrix models,
this identifies with a {\it classical}\/ limit for the operators involved,
the small parameter being $\hbar\equiv 1/N$.
Therefore in this limit,
the various operators $U$, $\tilde{U}$, $Z$ become classical variables
$u$, $v$, $z$,
up to some rescalings which will be detailed later.

What happens to the Toda equation? The large $N$ limit corresponds
to the dispersionless limit of the Toda hierarchy \refs{\TT,\TTb}.
In this limit, operators of discrete difference in the index $n$ 
(e.g.\ Eq.~\todab) can be replaced
with a differential operator $\der/\der \nu$ with $\nu\equiv n/N$.
Furthermore, the logarithm of the tau-function $\tau_N$ is required to be
of order $N^2$ (cf our Eq.~\defF),
with a smooth $1/N$ expansion,\foot{The
assumption of smoothness is less innocent than it seems; 
it is the equivalent of
the ``single cut'' hypothesis in matrix models (cf a similar
remark in \PZJc\ the context of 1D Toda). However in formal
power series around $t_q=\tilde{t}_q=0$ this hypothesis is
certainly valid.}
which leads to some simplifications in the differential equations.
We shall only write down a few examples of these dispersionless equations;
we refer the reader to e.g.\ \refs{\TT,\TTb,\Kri}
for a complete and more rigorous description of the dispersionless
hierarchy. Let us simply mention that, as mentioned above, we are dealing
with a classical limit, that is
the commutators become of order $\hbar\equiv 1/N$, and their leading
term defines a Poisson bracket. Equations \lax{}--\se\ can be rewritten
in the large $N$ limit, once rescaled,
by simply replacing commutators with Poisson brackets.
In particular, the string equation \se\ becomes:
\eqn\seb{
\{ u^{-1},v^{-1} \} = 1
}
(where now $u^{-1}$ and $v^{-1}$ are ordinary inverses since in the
large $N$ limit all operators are invertible).
It is worth noting that the same string equation occurs
in the interior Dirichlet boundary problem \MWZ. Also, it is identical
to the string equation of \Kos\ in which the radius has the unphysical
value $R=-1$.

\subsec{Scaling. Some examples}
In order to give some examples, we need to give some additional details on
the particular scaling of the variables that is appropriate for our model. 
First, we have for our operators $U$, $\tilde{U}$, $Z$: 
(cf the rescaling from Eq.~\twoMM\ to Eq.~\twoMMb):
\eqn\scal{
U\sim N^{-1/2} u\qquad \tilde{U} \sim N^{-1/2} v\qquad Z\sim z
}
where $u$, $v$, $z$ have a finite limit when $N\to\infty$
and are related by:
\eqna\expub
$$\eqalignno{
u&=r\, z+\sum_{k=0}^\infty u_k\, z^{-k}&\expub{a}\cr
v&=r\, z^{-1}+\sum_{k=0}^\infty v_k\, z^k&\expub{b}\cr
}$$
in terms of appropriately rescaled coefficients (in particular $r_N\sim
r N^{-1/2}$). We have already derived the rescaling of the moments,
cf Eq.~\iffb.

In what follows, we shall freely change from original to rescaled variables,
using the same letters by abuse of notation. In most circumstances this
does not cause any confusion.

For example, let us see what happens to the KP equation \KP. In the large $N$
limit, performing all the rescalings one term drops out and we find
\eqn\dKP{
2 {\der^2 \chi\over\der \theta_2^2}
+{\der\over\der \theta_1} \left(-2 {\der \chi\over\der \theta_3}
+\chi {\der \chi\over \der \theta_1}\right)=0
}
which is just the dispersionless KP equation (or Khokhlov--Zabolotskaia
equation).

Let now see what happens to
the Toda lattice equation \toda\ in the large $N$ limit. 
Here, it is important
to notice that an extra ingredient is required if one considers
equations in which variations with respect to $n$ appear.
It is our scaling hypothesis, based on the fact that we consider
only the ``planar'' large $N$ limit (first term in the $1/N$ expansion)
and on the scaling of the $t_q$ and $\tilde{t}_q$.
Considering Eq.~\defF, we write that:
\eqn\scalZ{
\tau_n = c_n\, \E{n^2 F[n^{-q/2-1}qt_q,n^{-q/2-1}q\tilde{t}_q]+O(1)}
}
The constant $c_n$ can be determined by setting $t_q=\tilde{t}_q=0$ in
Eq.~\detform; one easily finds $c_n=(\prod_{i=0}^{n-1} i!)^{-1}$.

More explicitly, if one introduces the dispersionless
tau function $F[\nu,\theta_q,\tilde{\theta}_q]=
\lim_{N\to\infty} ({1\over N^2}\log \tau_n+{1\over2} \nu^2 \log N-{3\over4})$ 
with $\nu=n/N$, then we have the scaling
\eqn\scalZb{
F[\nu,\theta_q,\tilde{\theta}_q]=-{1\over2}\nu^2 \log\nu + {3\over4}(\nu^2-1)
+\nu^2 F[\theta_q\nu^{-q/2-1},\tilde{\theta}_q\nu^{-q/2-1}]
}
where our old function $F[\theta_q,\tilde{\theta}_q]$ is recovered
by setting $\nu=1$.

We now rewrite the Toda equation~\toda:
\eqn\dtoda{
\E{{\der^2\over\der \nu^2} F}
={\der^2\over\der \theta_1\der\tilde{\theta}_1}F
}
which is just the dispersionless Toda equation (integrated twice);
then apply the scaling \scalZb.
The final result is the following partial differential equation for $F$:
\eqnn\scaltoda
$$\eqalignno{
&\exp\Big(2F+\sum_{q\ge 1}(\frac{q}{2}+1)(\frac{q}{2}-2)(\theta_q \der_q F+\tilde{\theta}_q \tilde{\der}_q F)\cr
&+\sum_{q,r\ge 1}(\frac{q}{2}+1)(\frac{r}{2}+1)(\theta_q\theta_r \der_q\der_r F+2 \theta_q\tilde{\theta}_r \der_q\tilde{\der}_rF
+\tilde{\theta}_q\tilde{\theta}_r \tilde{\der}_q\tilde{\der}_r F)\Big)
=\der_1\tilde{\der}_1 F&\scaltoda\cr
}
$$
with $\der_q\equiv\der/\der\theta_q$ and $\tilde{\der}_q\equiv
\der/\der\tilde{\theta}_q$. This is in general a fairly complicated equation.

\noindent{\it Application.} Assume that all $\theta_q$ and $\tilde{\theta}_q$
are zero except $\theta_1$ and $\tilde{\theta}_1$. 
This greatly simplifies Eq.~\scaltoda.
Note furthermore that by obvious homogeneity
property, $F$ only depends on the product $x\equiv \theta_1 \tilde{\theta}_1$.
Rewriting Eq.~\scaltoda\ for $F(x)$ leads to an ordinary differential equation:
\eqn\scaltodab{
\E{2F+9x^2 F''}=F' + x F''
}
with the initial condition $F(0)=0$,
which can be solved: if $\xi=x\,\d F/\d x$, then $\xi$ is the solution
of the third degree equation
\eqn\todasol{
16 \xi^3+ 8 \xi^2 + (1-36x) \xi + x (27 x-1)=0
}
that vanishes at $x=0$. We have the following expansion:
\eqn\todasolb{
\xi=\sum_{n=0}^\infty x^{n+1}
{(3n)!\, 2^n\over(n + 1)! \,(2n + 1)!}
}
Curiously, the number ${(3n)! 2^n\over (n+1)!(2n+1)!}$ also appears
in the context of the enumeration of planar maps;\foot{The author
would like to thank J.-B.~Zuber for pointing this out to him.}
it counts trivalent rooted maps with $2n$ nodes \Tut.
This suggests that a direct combinatorial proof of this result
might be possible. As a corollary, the HCIZ integral with $\theta_1$,
$\tilde{\theta}_1\ne 0$ is in the universality class of pure 2D quantum gravity:
the singularity of $\xi(x)$ closest to the origin
is of the form $\xi=(x-x_c)^{3/2}+\cdots$
(with $x_c=2/27$, $\xi_c=1/12$).

More generally, if one turns on a {\it finite}\/
number of $\theta_q$ and $\tilde{\theta}_q$,
one can show using the formalism developed in this section that
$F$ can be expressed in terms of the solution of an algebraic equation.
This is because the string equation implies that 
$a=1/u$ and $b=1/v$ are polynomials in $z^{-1}$, $z$:
\eqn\finwid{
a=\sum_{q=1}^{\tilde{\ell}+1} \alpha_q z^{-q}\qquad b=\sum_{q=1}^{\ell+1} \beta_q z^q
}
where $\ell=\max\{q|\theta_q\ne 0\}$, 
$\tilde{\ell}=\max\{q|\tilde{\theta}_q\ne 0\}$,
and as will be shown explicitly in the sections below (cf Eq.~\firstc),
the knowledge of $a$ and $b$ allows to compute derivatives of the
free energy.\foot{Note that since $a$, $b$ 
have a rational parameterization, the curve $(a,b)$ has genus zero.}

For example, if only $\theta_1$, $\theta_2$, $\cdots$, $\theta_\ell$, 
$\tilde{\theta}_1$
are non-zero, by homogeneity one can set $\tilde{\theta}_1=1$; then,
$\chi=2 {\der^2 F\over\der \theta_1^2}\equiv 2\psi^4$, 
and $r^2={\der^2 F\over\der\theta_1\der\tilde{\theta}_1}\equiv\psi$
are given by the degree $2\ell+1$ equation
\eqn\biguglysol{
-1+\psi+\sum_{q=1}^\ell (-1)^q {(2q)!\over (q!)^2} \theta_q \psi^{2q+1}=
-1 + \psi - 2 \theta_1 \psi ^3 + 6 \theta_2 \psi^5
-20 \theta_3 \psi^7+\cdots=0
}
with the initial condition $\psi=1$ for all $\theta_q$ zero.
One can check that if $\ell\ge 3$, $\chi=2\psi^4$ 
does satisfy the partial differential equation \dKP.

It is expected that by turning on an arbitrary number of $\theta_q$ and
$\tilde{\theta}_q$, one should be able to recover the critical behavior
(i.e.\ singularities of $F$ and its derivatives) of all $c<1$ minimal models
coupled to gravity.

\subsec{Derivatives of $F$ with respect to the $t_q$}
Our goal is now to express the first few
derivatives of $F$ with respect to
the $t_q$ and the $\tilde{t}_q$.
Let us define the differential operator
\eqn\defD{
\nabla(a)=\sum_{q\ge 1} {a^q\over q} {\der\over\der t_q}
}
In the bosonic language, this is the positive modes of the full
bosonic field $\phi(a)=\sum_{q\ge 1} {a^q\over q} {\der\over\der t_q}
+\log a-\sum_{q\ge 1} t_q a^{-q}$.
One could define similarly $\tilde{\nabla}(b)$. 

Our first task is to compute $\nabla(a)F$ (see also \refs{\PZJ,\PZJb} for
a ``saddle point'' approach to this question). We note that in terms of the
matrix model \twoMM,
$\nabla(a)F=-\lim_{N\to\infty}\left<\tr \log(1-a\U) \right>$, or
\eqn\first{
{\d\over\d a} \nabla(a) F = u^2 D(u)- u
}
where $D(u)\equiv \lim_{N\to\infty}
\left<\tr {1\over u-\U}\right>$ is the resolvent of $\U$,
and with the identification $u\equiv 1/a$ (we recall that appropriate
scaling as $N\to\infty$ is implied). Next we use the following
standard identity:
\eqn\pol{
p_N(u)=h_N^{-1} \left< \det(u-\U) \right>
}
where the average is again over $N\times N$ matrices with the measure
given by \twoMM.
This expression of $p_N(u)$ can be proved by checking that
$\braket{q_n}{p_N}=\delta_{nN}$ using it.
Considering the logarithmic
derivative with respect to $u$ of Eq.~\pol\ in
the large $N$ limit, we see that $D(u)$ is simply the operation of derivative
with respect to $u=1/a$ in the basis of the $p_n(u)$. This operator was
already introduced in the discussion of the string equation; Eq.~\preseb\
becomes, in the large $N$ limit,
\eqn\firstb{
D(u)=u^{-2} (v^{-1} - G(u)+u)
}
where we recall that $G(u)=u+\sum_{q\ge 1} q t_q u^{q+1}$ is
the resolvent of the original matrix $\A$. Combining Eqs.~\first\ 
and \firstb\ leads
to
\eqn\firstc{
{\d\over\d a}\nabla(a)F=b(a)-G(a)
}
where $b=1/v$, $a=1/u$ and $u$ and $v$ related by Eqs.~\expub{}, i.e.\ $b(a)$
is obtained by composing the series $b(z)=1/v(z)$ and the inverse series
$z(a)$ of $a(z)$.
Note that $b(a)-G(a)$ is simply the non-negative part of the Laurent expansion
of $b(a)={\d\over\d a} \phi(a)$ in powers of $a$.

Note that to derive the identity \firstc, we had to use the string equation
under the form \preseb. In contrast, the second derivative $\nabla(a_1)\nabla(a_2)F$
is ``universal'' i.e.\ in our context it means that it is true for any solution
of the dispersionless Toda hierarchy regardless of the string equation.
Derivation of $\nabla(a_1)\nabla(a_2)F$ requires two identities.
We shall prove the first one, which is an expression for $z$. 
We note that since $Z$ is the shift operator,
one should have in the large $N$ limit $z=P_{n+1}(u)/P_n(u)$, that is taking
the logarithm and using Eq.~\pol,
\eqn\idz{
\log z = \log u - \log r -\sum_{q\ge 1} {u^{-q}\over q} {\der^2\over\der\nu\der t_q}F
}
where it is recalled that the dependence on $\nu$ of $F$ is given by Eq.~\scalZb.

The second relation 
we use the dispersionless limit of the Fay identity \TTb, which 
we shall not prove here and write under the form (see \WZ):
\eqn\fay{
(u_1-u_2)\E{\sum_{q,r\ge 1} {u_1^{-q}\over q} {u_2^{-r}\over r}
{\der^2\over\der t_q \der t_r} F}
= u_1\, \E{-\sum_{q\ge 1} u_1^{-q}{\der^2\over\der\nu\der t_q} F}
- u_2\, \E{-\sum_{q\ge 1} u_2^{-q}{\der^2\over\der\nu\der t_q} F}
}
Combining Eqs.~\idz\ and \fay, using the variables $a_1=1/u_1$,
$a_2=1/u_2$ and the power series inverse $z(a)$ of $a(z)$ leads to
\eqn\twob{
\nabla(a_1)\nabla(a_2) F = \log {z(a_1)-z(a_2)\over 1/a_1-1/a_2}+\log r
}
(cf also a similar formula in \DKK).

One can prove a formula for mixed derivatives:
$\nabla(a)\tilde{\nabla}(b)F=-\log(1-z(b)/z(a))$,
but it will not be needed here.

Formulae for higher numbers of derivatives become more and more complicated,
(see in particular ``residue formulae''
for three derivatives \Kri), and we shall not write
them down here.

\newsec{Expansion around $0$ of the HCIZ integral}
We now come to the main goal of this paper: to apply the equations above
to the expansion of the large $N$ HCIZ free energy $F[\theta_q,\tilde{\theta}_q]$
in powers of $\theta_q$ and $\tilde{\theta}_q$. We shall consider the problem
in an asymmetric way: we shall apply the operator $\nabla(a)$ defined 
in the previous section to the free energy, 
that is consider derivatives with respect
to the $\theta_q$, and then set $\theta_q=0$ and see how the result depends
on the $\tilde{\theta}_q$. The identities obtained above involve
the functions $b(a)=1/v(a=1/u)$ and $z(a)$; 
let us see how to compute these functions when all $t_q$ are zero.

If we set $t_q=0$ in def.~\bil\ of the bilinear form, we see that no positive
powers of $u$ are generated, so that 
$m_{ij}=\braket{v^i}{u^j}$ satisfies
\eqn\trim{
m_{ij}=\cases{
0& $j<i$\cr
{1\over j!}\,\tilde{s}_{j-i}& $j\ge i$}
}
where we have defined $\tilde{s}_q$ by 
$\varphi(v)\equiv \exp\big(\sum_{q\ge 1} \tilde{t}_q
v^q\big)=\sum_{q\ge 0} \tilde{s}_q v^q$; they are Schur functions
corresponding to single row Young diagrams.

Let us first use the fact that $M$ is upper triangular.
Since $M=\tilde{P} P^{-1}$ with $P$ upper triangular and $\tilde{P}$ lower
triangular, $\tilde{P}$ is diagonal i.e.\ $q_n(v)=\sqrt{n!}\, v^n$.
The normalization $h_n=(n!)^{-1/2}$ implies that $r_n=n^{-1/2}$, 
that is after
rescaling $r=1$. Finally, we conclude from the definition \defuv\ of $V$
that all the coefficients $v_k$ in its expansion (or of its transpose
$\tilde{U}$, Eq.~\expu) are zero, or in the large $N$ limit
\eqn\zero{
v=z^{-1}
}
(note that this is consistent with Eq.~\finwid).
We have now reduced the problem to a single function $b(a)
\equiv 1/v(a)=z(a)$; its calculation was performed in \PZJb\ using
a saddle point method, but it is particularly simple to rederive it
in our framework. We start again from Eq.~\trim\ 
and notice that one
can easily invert this matrix, which yields the coefficients
of $P=M^{-1} \tilde{P}$. If we introduce the $\tilde{e}_q$
defined by $\varphi(v)^{-1}\equiv \exp\big(-\sum_{q\ge 1} \tilde{t}_q v^q\big)
=\sum_{q\ge 0} \tilde{e}_q v^q$ (they are $(-1)^q$ times the
Schur functions corresponding to one-column Young diagrams),
then expansion of the identity $\varphi(v)\varphi(v)^{-1}=1$ leads to
\eqn\yay{
p_{ij}=\cases{
0& $j<i$\cr
{i!\over\sqrt{j!}}\, \tilde{e}_{j-i}& $j\ge i$
}
}

We now define the matrix $A=(a_{ij})$ by
$A=P^{-1} Z^T P$, which according
to defs.~\defuv\ is a left inverse of $U$. We have
\eqnn\yayb
$$\eqalignno{
a_{ik}&= \sqrt{i!\over k!} \sum_{j,\, i\le j\le k-1}
{1\over j!}\tilde{s}_{j-i} (j+1)! \tilde{e}_{k-1-j}\cr
&=\sqrt{i!\over k!} \sum_{q=0}^{k-1-i}
(q+i+1) \tilde{s}_q \tilde{e}_{k-1-i-q}&\yayb\cr
}
$$
This time we recognize the expansion of
$\varphi'(v)\varphi(v)^{-1}=\sum_{n\ge 0} v^{n-1} \sum_{q=0}^n q \tilde{s}_q
\tilde{e}_{n-q}$; but we also have $\varphi'(v)\varphi(v)^{-1}=\sum_{n\ge 1}
n \tilde{t}_n v^{n-1}$, so that
\eqn\yayc{
a_{ik}=\cases{
0& $k<i+1$\cr
\sqrt{k}& $k=i+1$\cr
\sqrt{i!\over k!}\, (k-i-1) \tilde{t}_{k-i-1} & $k> i+1$
}
}
This is an exact expression at finite $N$.
In the large $N$ limit this can be simply rewritten, performing
the appropriate rescalings and remembering that $z=b$,
\eqn\yayd{
a={1\over b}+\sum_{q\ge 1}{\tilde{\theta}_q\over b^{q+1}}
}
We recognize the resolvent of the original
matrix $\B$ (Eq.~\defres): $a(b)=\tilde{G}(b)$ (again this is consistent
with the ``dual'' equation of \firstc, ${\d\over\d b} \tilde{\nabla}(b)
F=a(b)-\tilde{G}(b)$). 
Finally $b(a=1/u)$ is the functional inverse of $a(b)$;
it is well-known that $b(a)$ is nothing but the generating
series of {\it free cumulants}\/ $\tilde{m}_q$ of $\B$ \Spe\ 
(cf also the expression of {\it connected}\/
correlation functions in matrix models, as in Eq.~(31) of \BIPZ):
\eqna\yaye
$$\eqalignno{
b(a)&={1\over a}+\sum_{q=1}^\infty \tilde{m}_q a^{q-1}&\yaye{a}\cr
\tilde{m}_q&=-\sum_{\alpha_1,\ldots,\alpha_q\ge 0\atop\sum_i i\alpha_i=q} 
{(q+\sum_i \alpha_i-2)!\over (q-1)!} 
\prod_i {(-\tilde{\theta}_i)^{\alpha_i}\over\alpha_i!}&\yaye{b}\cr
}
$$

We can now apply the formulae obtained in previous section.
We recall that 
$\nabla(a)\equiv\sum_{q\ge 1} {a^q\over q} {\der\over\der t_q}$. First
we rewrite Eq.~\firstc\ for $t_q=0$:
\eqn\firstz{
{\d\over\d a} \nabla(a) F
 = b(a) - {1\over a}=\sum_{q=1}^\infty \tilde{m}_q a^{q-1}
}
This equation can be found in \PZJb, but it
can already be extracted from \IZ; see also \Co\ for a combinatorial proof.

The second derivatives are given by Eq.~\twob\ with $z(a)=b(a)$:
\eqnn\secz
$$\eqalignno{
\nabla(a_1)\nabla(a_2) F &= \log {b(a_1)-b(a_2)\over 1/a_1-1/a_2}\cr
&=\log\left(1-\sum_{q=2}^\infty \tilde{m}_q 
\sum_{k=1}^{q-1} a_1^k a_2^{q-k}\right)&\secz\cr
}$$
Since the expression is in fact known for all $t_q$, one can differentiate once
more and obtain the expression for the third derivatives at $t_q=0$
with little extra work. First
one computes the expression of $b(z)$ at first order in $t_q$ via Eq.~\lax{b};
one obtains after inversion
\eqn\bom{
r\,z=b+\sum_{q\ge 1} t_q \sum_{k\ge 0} c_{qk} b^{k+1}+\cdots\qquad t_q\to 0
}
where the coefficients $c_{qk}$ are best expressed using their
generating function
\eqn\coef{
\sum_{q\ge 1} {a^q\over q} c_{qk}={\d b(a)\over \d a} b(a)^{-k-2}
}
Secondly one uses Eqs.~\firstc\ and \twob\ to deduce
\eqn\ded{
\nabla(a_2) b(a_1)=
{\der\over \der a_1} \log(z(a_1)-z(a_2))
}
Acting with $\nabla(a_3)$ on Eq.~\twob, setting $t_q=0$,
using identities \bom--\ded\ and
performing simple manipulations results in
\eqnn\thirdz
$$\eqalignno{
\nabla(a_1)\nabla(a_2)\nabla(a_3)F&=
{\d b(a_1)/\d a_1\over (b(a_1)-b(a_2))(b(a_1)-b(a_3))}
+{\d b(a_2)/\d a_2\over (b(a_2)-b(a_1))(b(a_2)-b(a_3))}\cr
&+{\d b(a_3)/\d a_3\over (b(a_3)-b(a_1))(b(a_3)-b(a_2))}
+1&\thirdz}
$$
One can go to higher derivatives by further differentiation, but 
most likely, the expressions become more and more cumbersome.

\newsec{Summary of results}
In this paper we have investigated the expansion of the large $N$
HCIZ integral free energy in terms of powers
of the moments of the matrices. We have shown that this free energy is
the dispersionless tau function for the dispersionless
2D Toda hierarchy of partial
differential equations, so that coefficients of the expansion satisfy
an infinite set of non-linear relations.
Also, we have seen that the solution
of the Toda hierarchy is selected by a non-trivial string equation
$\{ u^{-1}, v^{-1} \}=1$. The following explicit results have been found:
\smallskip
$\diamond$ First, one can compute some infinite series of
coefficients of the expansion using these differential equations.
The example worked out in section 4 is the set of coefficients of
the form $\theta_1^k \tilde{\theta}_1^k$; if $x=\theta_1 \tilde{\theta}_1$,
then $\xi=x\,\d F/\d x$ was found to satisfy the equation
$16 \xi^3+ 8 \xi^2 + (1-36x) \xi + x (27 x-1)=0$, that is
\eqn\Fxp{
F=\theta_1\tilde{\theta}_1 + {1\over2}(\theta_1\tilde{\theta}_1)^2 + {4\over3} (\theta_1\tilde{\theta}_1)^3 +6 (\theta_1\tilde{\theta}_1)^4 + {176\over5} (\theta_1\tilde{\theta}_1)^5+\cdots
}

%
More generally, one can compute generating series of
coefficients involving only a finite number of distinct 
$\theta_q$ and $\tilde{\theta}_q$ as solutions of algebraic equations;
a more complicated example was given without proof, cf Eq.~\biguglysol.
In principle, this allows to compute any given coefficient required,
although it does not give a closed expression for the whole function of
the infinite set of $\theta_q$ and $\tilde{\theta}_q$.

\smallskip
$\diamond$ Secondly, one can obtain {\it all}\/ the coefficients (as
functions of the $\tilde{\theta}_q$) of
the first few orders of the expansion in products of $\theta_q$: 
we rewrite their generating functions here (Eqs.~\firstz, \secz\ and \thirdz):
\eqna\summ
$$\eqalignno{
{\d\over\d a} \nabla(a) F
 &= b(a) - {1\over a}&\summ{.1}\cr
\nabla(a_1)\nabla(a_2) F &= \log {b(a_1)-b(a_2)\over 1/a_1-1/a_2}&\summ{.2}\cr
\nabla(a_1)\nabla(a_2)\nabla(a_3)F&=
{\d b(a_1)/\d a_1\over (b(a_1)-b(a_2))(b(a_1)-b(a_3))}
+{\d b(a_2)/\d a_2\over (b(a_2)-b(a_1))(b(a_2)-b(a_3))}\cr
&+{\d b(a_3)/\d a_3\over (b(a_3)-b(a_1))(b(a_3)-b(a_2))}
+1&\summ{.3}\cr
}$$
where in the rescaled variables,
$\nabla(a)=\sum_{q\ge 1} a^q {\der\over\der \theta_q}$. 
The expansion is given in terms of the function 
$b(a)=1/a+\sum_{q\ge1}\tilde{m}_q a^{q-1}$.
The free cumulants $\tilde{m}_q$
can be reexpanded themselves in powers of the $\tilde{\theta}_q$:
$\tilde{m}_q=\tilde{\theta}_q-{q\over 2!} \sum_{i,j\ge 1\atop i+j=q} 
\tilde{\theta}_i \tilde{\theta}_j
+{q(q+1)\over 3!}\sum_{i,j,k\ge 1\atop i+j+k=q}
\tilde{\theta}_i \tilde{\theta}_j \tilde{\theta}_k+
\cdots$,
so that we find for example
\goodbreak
\eqnn\finalyay
$$\eqalignno{
F&=\sum_{q\ge 1} {1\over q} \theta_q \tilde{\theta}_q\cr
&-{1\over2}\sum_{\scriptstyle q,r_1,r_2\ge 1\atop\scriptstyle r_1+r_2=q}
\theta_q \tilde{\theta}_{r_1}\tilde{\theta}_{r_2}\cr
&-{1\over2}\sum_{\scriptstyle r,q_1,q_2\ge 1\atop\scriptstyle q_1+q_2=r}
\theta_{q_1}\theta_{q_2} \tilde{\theta}_r\cr
&+{1\over4}\sum_{\scriptstyle q_1,q_2,r_1,r_2\ge 1\atop\scriptstyle q_1+q_2=r_1+r_2}
(q_1+q_2-\min(q_1,q_2,r_1,r_2)+1)\theta_{q_1}\theta_{q_2} \tilde{\theta}_{r_1}
\tilde{\theta}_{r_2}\cr
&+{1\over6}\sum_{\scriptstyle q,r_1,r_2,r_3\ge 1\atop\scriptstyle q=r_1+r_2+r_3}
(q+1)\theta_q \tilde{\theta}_{r_1}\tilde{\theta}_{r_2}\tilde{\theta}_{r_3}\cr
&+{1\over6}\sum_{\scriptstyle r,q_1,q_2,q_3\ge 1\atop\scriptstyle r=q_1+q_2+q_3}
(r+1)\theta_{q_1}\theta_{q_2}\theta_{q_3} \tilde{\theta}_r\cr
&+\cdots&\finalyay
}$$
with the appropriate symmetry of interchange of the $\theta_q$ and
$\tilde{\theta}_q$.

\bigskip\centerline{\bf Acknowledgements}
The author would like to thank P.~Biane, B.~Collins,
B.~Eynard and especially I.~Kostov and J.-B.~Zuber for discussions.

\listrefs
\bye